\documentclass{pnastwo-modified}

\advance\voffset -.5in 

\usepackage{graphicx}

\usepackage{amssymb,amsfonts,amsmath}
\usepackage{pslatex}

\begin{document}



\title{Neutral networks of genotypes: Evolution behind the curtain}
\def\theurl{Neutral networks of genotypes: Evolution behind the curtain}





\author{Susanna C. Manrubia%
\affil{1}{Centro de Astrobiolog\'{\i}a, CSIC-INTA, Ctra.~de Ajalvir km.~4,
28850 Torrej\'on de Ardoz, Madrid, Spain}
\and
Jos\'e A. Cuesta%
\affil{2}{Grupo Interdisciplinar de Sistemas Complejos (GISC),
Dept.~de Matem\'aticas, Universidad Carlos III de Madrid,
Avda.~de la Universidad 30, 28911, Legan\'es, Madrid, Spain}
}

\def\thefootlineauthor{Manrubia and Cuesta}

\contributor{}

\maketitle

\begin{article}

\begin{abstract}
Our understanding of the evolutionary process has gone a long way since the publication,
150 years ago, of ``On the origin of species'' by Charles R. Darwin. The XXth Century
witnessed great efforts to embrace replication, mutation, and selection within the
framework of a formal theory, able eventually to predict the dynamics and fate of
evolving populations. However, a large body of empirical evidence collected over the
last decades strongly suggests that some of the assumptions of those classical models
necessitate a deep revision. The viability of organisms is not dependent on a unique and
optimal genotype. The discovery of huge sets of genotypes (or neutral networks) yielding
the same phenotype ---in the last term the same organism---, reveals that, most likely,
very different functional solutions can be found, accessed and fixed in a population
through a low-cost exploration of the space of genomes.
The `evolution behind the curtain' may be the answer to some of the current puzzles
that evolutionary theory faces, like the fast speciation process that
is observed in the fossil record after very long stasis periods.
\end{abstract}

\keywords{Neutral network | genotype-phenotype map | redundancy | adaptation |
          fitness landscape}





\section{Introduction}

\begin{figure}[t]
\centerline{\includegraphics[scale=0.44,clip=]{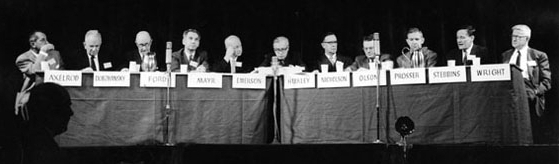}}
\caption{``The Evolution of Life'', November 25, 1959. Darwin Centennial Celebration.
From left to right: Daniel I.\ Axelrod, Theodosius Dobzhansky, Edmund B.\ Ford, Ernst
Mayr, Alfred E.\ Emerson, Julian Huxley, Alexander J.\ Nicholson, Everett C.\ Olson, 
Clifford Ladd Prosser, George Ledyard Stebbins, and Sewall Wright.
\label{fig:conference} 
}
\end{figure}

The first name that comes to our minds when we hear the word `evolution' is
Darwin. No doubt that Charles Robert Darwin's \emph{On the Origin of
Species}~\cite{Dar59}, together with the sequels that he also published
\emph{(The Descent of Man, The Expression of Emotions in Man and Animals\dots)},
form the cornerstone of our current understanding of the most fundamental
process of life. Nevertheless, Darwin neither discovered evolution himself, nor
was he the only one to propose the mechanism of natural selection to explain
the evolution of species. At Darwin's time, the fact that species evolved was common
knowledge.\footnote{Jean-Baptiste Lamarck's work,
appeared in 1802, is considered the first ---though
incorrect--- published theory of evolution. Lamarck's ideas were anticipated by
Erasmus Darwin, one of Darwin's grand-fathers, and even earlier by
Maupertuis, who envisioned a genetic inheritance of characters, entertained the
idea that new species arise as mutant individuals, and even considered the
elimination of deficient mutants, thus suggesting some kind of natural
selection \cite{May85}.}
On the other hand, Alfred Russel Wallace published, simultaneously with Darwin,
a theory of evolution based on what we currently know as natural selection, the
same key idea put forward in ``The Origin''. Then, why is Darwin's work so 
fundamental for the current theory of evolution? To understand the depth
of his contribution, one must read ``The Origin'' ---just an abstract, in his
words, of the work he intended to publish two or three years later \cite{Dar59}.
He deserves the credit for this theory
because of both the overwhelming accumulation of empirical data he presented
and the clear explanations that his theory offered to many
different ---and at the time independent--- observations: geographical diversity,
artificial selection, coevolution of plants and insects, appearance of complex
organs, instincts in man and animals\dots{} He gave a unified view of the 
complexity of life by means of a unique universal mechanism. Evolution by
natural selection was endowed with a creative power far beyond
what Darwin's predecessors, or even Wallace, had ever proposed \cite{Gou02}.
It is for this reason that there was a centennial celebration
of the publication of this fundamental
book (see Fig.~\ref{fig:conference}) and the motive of last year's
sesquicentennial celebration, in the internationally proclaimed Darwin year.

However, Darwin's theory was incomplete. All throughout ``The Origin'', 
Darwin bumps once and again into the same problem: the mechanism
of inheritance. At Darwin's time the standard theory of inheritance 
in sexual organisms assumed that individuals roughly inherited an average
of their parent's traits. Sir Francis
Galton, one of Darwin's cousins, discovered the statistical phenomenon of
regression towards the mean \cite{Gal86}, according to which traits that
deviate from the mean of a population revert to this value as they breed,
in a few generations. This problem permeates 
his work and forces Darwin
to resort to the isolation of populations in order to explain the appearance
and maintenance of new species. It was unfortunate that Darwin was not
aware of Mendel's discovery of the laws of inheritance, published almost
simultaneously with ``The Origin'' in an obscure Austrian journal \cite{Men66}.
Mendel laws would have solved many of Darwin's problems with the sustainment
of diversity. In fact, the rediscovery of these laws by de Vries, Correns and
von Tschermak in 1900 triggered a big deal of research, both theoretical and
experimental, which led, by the middle of the XXth century, to the so-called
``modern synthesis'' \cite{Hux42}. This revision of Darwinism
can be considered
as a true scientific theory in the sense that it is based on population
genetics, a \emph{quantitative} formulation of the theory of evolution by
natural selection under the mechanisms of genetic inheritance.

\section{The current paradigm: population genetics}

Population genetics is the creation of a group of statisticians among whom
we find some of the big names of evolutionary theory: Fisher, Haldane,
Wright, and later Kimura. The focus of this theory is to determine the
fate of a population whose individuals reproduce with variability and struggle
for survival in an environment which discriminates their traits, favoring
ones over others. More precisely, population genetics assumes
that populations live in a more or less exhausted environment which maintains
the amount of individuals almost constant along generations. Individuals
breed and their offspring inherit their traits according to genetic laws.
Different traits have different survival probabilities, and the action of 
chance upon this biased set decides
who dies with no descent and who survives and reproduces, and, among the latter,
the number of offspring of each individual. New traits appear randomly,
at a very low rate, through mutations of existing genes. 
From this point of view, evolution is, to a large extent, a result of the laws of 
probability, hence the intrinsic statistical nature of population genetics.

Population genetics stands as the first coherent and quantitative account of
the theory of evolution, and still today provides the paradigm that scientists
have in mind when thinking about evolution. The picture it draws is that of
a population of entities which \emph{replicate} at a rate that depends
on \emph{selection} pressures, i.e.\ a measure of how adapted are their traits
to the environment. New traits appear at a very low rate through \emph{mutations.}
The process is random and therefore subject to historical contingency, which
translates into another feature exhibited by the evolution of populations:
\emph{genetic drift}, or sampling noise. By this we mean the fact that even
for a population with two traits replicating at the same rate ---i.e.\ having
the same \emph{fitness}--- and represented fifty-fifty, the ratio of the two
traits will deviate from this equal ratio in the next generation. This process
is especially important in small populations (for instance, in evolutionary
bottlenecks), but it has always been considered a secondary effect in large
populations. The paradigm yielded by population genetics
has been very successful not only in Biology, but also in other
disciplines which have borrowed it to explain related phenomena. Economics,
Sociology, Linguistics, or Computer Science are a few examples of areas where
evolution as a result of the combined effect of replication, selection, and
mutation, has provided a new framework to understand 
collective dynamics or to devise applications to solve existing problems.

But population genetics also makes several implicit assumptions which have
basically remained unquestioned and have thus become part of the standard
thinking in this discipline. Explicit models in population genetics make
use of a metaphor introduced by Wright: the fitness landscape. In brief,
it is assumed that fitness is uniquely determined once the genotype and the
environment are given, so if the environment remains unchanged, the fitness
landscape becomes a mapping from genotype to the mean replication rate
(interpreted as fitness) of the
individuals carrying that genotype. Evolution is then the movement through
that fitness landscape. But what does it move? This is the first implicit
assumption of population genetics: evolution moves
the population as a whole. The mutation
rate is considered so low that 
a mutation causing a new allele gets fixed in the population
before the next mutation occurs and introduces a new allele into play. Thus
evolution is the movement of a homogeneous population throughout the fitness
landscape. This implicit assumption is made explicit in several works aimed
at describing the evolution of populations with the language of Statistical
Mechanics \cite{Sel05,Bar09}. A second implicit assumption shows up when
examining the basic models of population genetics. Fisher's Fujiyama landscape
assumes, for instance, that there is an optimum genotype for which fitness
is maximal, and any deviation from that genotype by point mutations only
degrades that fitness, the more the larger the distance in configurational
space (genotype distance is 
usually measured in terms of Hamming distance,
i.e.\ the number of positions in which two sequences differ). Wright's
rugged landscapes are thought of as hilly landscapes, with many mountains
and valleys, tops being fitness maxima, again located at specific genotypes.
Many theoretical models like Muller's ratchet \cite{Mul32} or Eigen's
quasispecies
\cite{Eig71}, which have been very influential in our current
evolutionary thinking, strongly rely on this optimum genotype assumption
of population genetics.

{\it Gradualism} is implicit in this evolutionary paradigm:
evolutionary changes occur only through the gradual, slow accumulation
of small changes caused by the very 
infrequent appearance of beneficial
mutations (most mutations are just deleterious).
Gradualism, an idea that Darwin took from Geology, is one
of the strong arguments of ``The Origin'' in justifying why we are not
able to see evolution at work. We cannot see it like we cannot see
mountains erosion, and yet we know it exists. But gradualism is also one
of the most controversial points of evolutionary theory because it
conflicts with the fossil record, where species are observed to remain
nearly unchanged for long stasis periods, only to be quickly (in
geological terms) replaced by new species (something that has been
termed \emph{punctuated equilibrium} \cite{Eld72}).

Gradualism is only the tip of the iceberg. Perhaps it is so because a case
can be made against it from the empirical evidence accumulated by more
than a century of paleontological research and from the accumulated
knowledge on non-parsimonious evolutionary mechanisms. Still, it is not the only
difficulty that the paradigm of population genetics faces, nor is it
the first one to show up. We will see immediately that the strongest
body of evidence against many of the assumptions underlying population
genetics comes from molecular biology. And it urgently calls for a
change of paradigm. This does not mean that population genetics is
wrong: on the contrary, the tools it provides are still valid. It is
only the picture it draws, more based on somehow prejudicial assumptions
and on misleading metaphors, that is essentially incorrect.

\section{The new paradigm: neutral evolution}

In 1968 Kimura surprised the scientific community with the argument that
most mutations in the genome of mammals have no effect on their phenotype
\cite{Kim68}: in other words, most mutations are \emph{neutral,} neither
beneficial nor deleterious. The argument goes as follows.
Comparative studies of some proteins indicate that in chains nearly
100 aminoacids long a substitution takes place, on average, every
$28$ million years. The typical length of a DNA chain in one of the
two sets of mammal chromosomes is about $4$ billion base pairs.
Every $3$ base pairs \emph{(codon)} code for an aminoacid and, because
of redundancy, only 80\% base
pair substitutions give rise to an aminoacid substitution in the
corresponding protein. Therefore there are $16$ million substitutions
in the whole genome every $28$ million years; in other words, approximately
a substitution every $2$ years! Kimura concluded that such an enormous
mutational load can only be tolerated if the great majority of mutations
are neutral.

Subsequent studies with different systems (we will see later the case of RNA
molecules) support this conclusion. 
At least at the molecular level, neutrality seems to be the rule,
rather than the exception, thus contradicting
the homogeneity assumption of population genetics. One could
argue that neutral mutations can simply be disregarded,
so that we can just focus
on those that do produce a phenotypic change in the individual.
This might be an appropriate description of what is going on
if the effect of mutations on phenotype, and
therefore on fitness, could be added up, as if genes were simple switches
of different traits that can be turned on and off by mutations
(unfortunately a widespread misconception of how genes work).
But things
are far less simple. It turns out that genes are involved in a complex
regulatory network in which the proteins codified by some genes 
activate or inhibit the coding of other proteins (even themselves),
so that the action of a single protein ---hence of a gene--- cannot be
disentangled from the action of very many others. In fact, there is
nearly no single trait in multicellular animals or plants which is not
the consequence of the combined effect of many genes acting together
in this complex way.

The phenotype is thus the effect of the genome as a whole, rather
than a `linear combination' of traits. Now, the accumulation of neutral
mutations motivates that
apparently similar individuals of the same
species bear genomes that may be very far apart from each other.
In this situation a new mutation may induce a big phenotypic
change in one of these individuals but not in others
because the net effect is as if the genome as a whole had
been modified in just one step (all previous mutations were silent).
This effect challenges the standard picture of gradualism
and makes a case for
punctuated equilibrium. Not only that: the idea that
there is an optimum genotype makes no sense under such a wide neutral
wandering in the space of sequences, and this, as we will see, questions
many commonly accepted models in population genetics.

\section{Variability and redundancy}

\begin{figure}
\centerline{\includegraphics[scale=0.3,clip=]{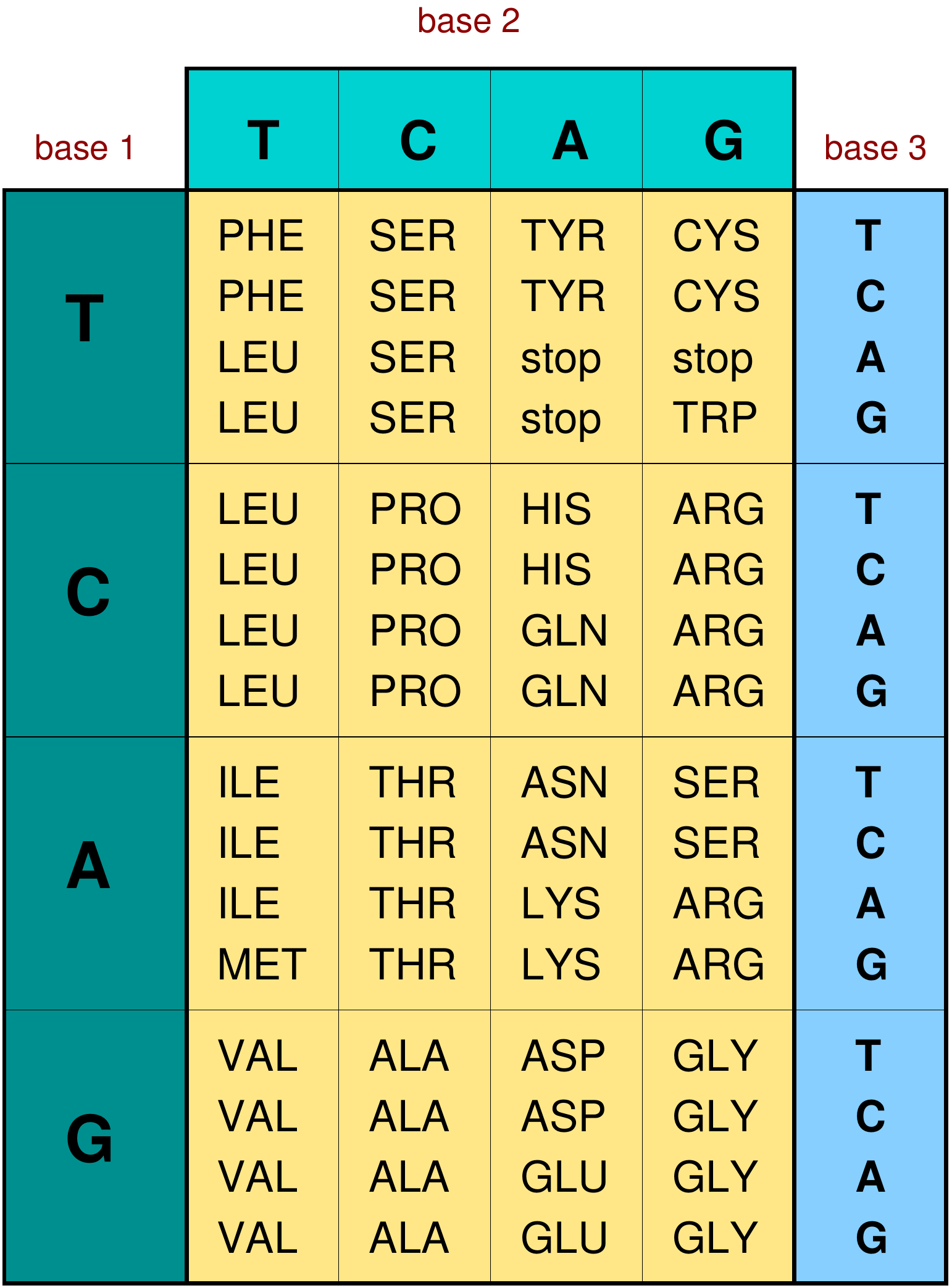}}
\caption{Genetic code. A, T, G, C stand for the four basis of DNA (Adenine,
Thymine, Guanine, and Cytosine). Transcription is carried out by
RNA chains, which are copies of one DNA strand with T replaced by U (Uracil).
PHE, LEU, ILE, etc., are abbreviations of the $20$ aminoacids (PHEnylalanine,
LEUcine, IsoLEucine, etc.). Codons labelled `stop' signal the end of 
transcription.
\label{FGC}
}
\end{figure}

Biology is extremely redundant, and it is so at all its levels of complexity.
We have just mentioned the redundancy of the genetic code.
Every codon codes for an aminoacid using an
almost universal code (see Fig.~\ref{FGC}). Setting aside three
`stop' codons (which mark the end of the gene), this implies that
61 codons code for only 20 aminoacids. Thus most aminoacids are coded
by two, four, or even six codons, 
so many base pair substitutions
in the DNA do not alter the coded protein. Proteins, in their turn,
fold in an almost rigid three-dimensional structure (the so-called
\emph{tertiary} structure). This folding is induced by the interaction
between the sequence of aminoacids conforming their \emph{primary} structure.
But not all aminoacids play the same role in folding the protein:
some of them are critical, in the sense that if they are replaced
by others the conformation of the protein changes, but most are 
nearly irrelevant, in the sense that their replacement leaves the
protein unchanged or nearly so. As the tertiary structure determines
the protein function, it turns out that many aminoacid substitutions
do not modify the structure, and thus have no biological effect. Proteins
then enter a complex regulatory or metabolic network in which they
interact with other proteins regulating their coding
or participating in metabolic pathways. But then again some of this
proteins may be replaced by other similar proteins with no major change
in the network function.

This extraordinary redundancy of biological systems makes them
very robust to change. This is the origin of neutrality. In order to understand
how much room for neutrality is there in biological systems and grasp
some of the effects induced by variability 
we will closely examine a relatively simple example to which a big deal
of research has been devoted in the last decades: RNA folding
\cite{Anc00,Fon02,Sch06}. An RNA molecule is a chain formed
by a sequence of the four nucleotides G, C, A, and U. Although it
can form double chains, as DNA does, RNA molecules
are usually single stranded. Nucleotides in an RNA sequence
tend to form pairs to minimize the free energy of the molecule. This so-called
secondary structure of RNA molecules determines to a large extent their chemical
functions, and as such has been often used as a crude representation of
the phenotype. 

\begin{figure}
\centerline{\includegraphics[scale=0.44,clip=]{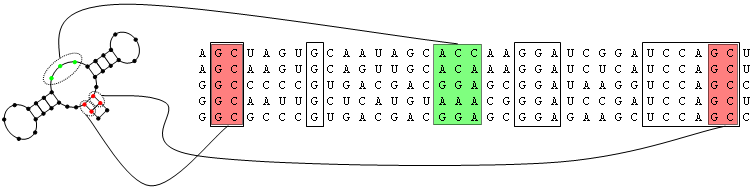}}
\caption{An example of an RNA secondary structure and a few of the sequences
that fold into that state as their configuration of minimum energy. We
highlight the only conserved regions in this example (nucleotides surrounded
by solid-line boxes), which typically correspond to nucleotides forming
pairs in the structure (a particular case shown in pink). Non-paired 
nucleotides form loops in the secondary structure, and are less conserved
on average than stacks (e.g.~the three positions forming the internal loop,
indicated in green in the sequences).
\label{F2} 
}
\end{figure}

An upper bound for the number $S(l)$ of sequences of length $l$ compatible
with a fixed secondary structure is $S(l)\propto l^{-3/2}b^l$~\cite{Sch94}, 
where $b$ is a constant that depends on geometric
constraints imposed on the secondary structure (e.g.\ the minimum number of
contiguous pairs in a stack). The calculation of $S(l)$ is done in a recursive
manner, summing over all possible modifications of a structure when its length
increases in one nucleotide. The resulting equations may be considered a 
generalization of Catalan and Motzkin numbers~\cite{Wat78}.
The values of $S(l)$ for moderate $l$ are
certainly huge: there are about $10^{28}$ sequences compatible with the
structure of a transfer RNA (which has length $l=76$), while the currently
known smallest functional RNAs, of length $l \approx 14$ \cite{And05}, could
in principle be obtained from more than $10^6$ different sequences.
Figure~\ref{F2} portrays a computational example of sequences folding into
the same secondary structure, of length $l=35$ in that case. Note that the
similarity between sequences may be very low, even if they share their folded
configuration: a random subsample of a population reveals that sequences
differ on average in 10 to 15 nucleotides, while differences up to 100\% are
possible. 

All this enormous variability that redundancy supports may have a measurable
effect: the equilibrium configuration of either large populations or of populations
evolving at a high enough mutation rate, is very heterogeneous. For the sake of 
illustration let us consider a population of size $N$ undergoing a mutation
rate $\mu$ per generation and per individual. To simplify, let us also assume
that all mutations are neutral. In this case, the time $t_g$ in number of
generations required for a mutation to spread to all individuals (or to disappear)
is proportional to the population size, $t_g \simeq 2 N$ \cite{Ewe04}.
Now, the number $M$ of mutants that appear in this characteristic time is
$M \simeq t_g \mu N = 2 N^2 \mu$. The conclusion is straight: if $M\sim 1$
the population will be homogeneous most of the time, but if $M\gg 1$ mutants 
appear at a rate faster than that at which mutations are fixed in the whole
population, so the statistical equilibrium will correspond to a heterogeneous
population. 

Heterogeneity is dynamically maintained not only in neutral characters, but also in 
features that affect fitness. There are abundant observations of suboptimal phenotypes 
that coexist with better adapted phenotypes. This is also a result of a high mutation 
rate that translates into non-zero transition probabilities between phenotypic classes. 
In other words, the existence of just one of the phenotypes generates all the others, 
which are mutually maintained at equilibrium. This type of organization is called a 
quasispecies. It was first introduced in a theoretical setting to describe the 
organization of macromolecules at prebiotic times \cite{Eig71}, and the concept was 
subsequently applied to viruses \cite{Dom78}. Actually, RNA viruses yield abundant 
examples of heterogeneity, both in sequences and in function. The common situation is 
that each genotype is unique in the population, differing in at least one nucleotide 
from any other. But the isolation of those genotypes and the subsequent generation of 
clonal populations that descend from each of them reveals 
a high variability in phenotypic properties (replication time or virulence, 
for instance), such that the population is a heterogeneous ensemble in genotype 
and phenotype~\cite{Dua94}.

\section{Distribution of phenotypes in genotype space}
\label{sec:V}

Genotype spaces are rather complex objects amenable to a deceptively simple
description. So complex and so simple that thinking of them may easily lead
to misleading images that 
misguide our intuition. Much of our difficulty
in understanding the dynamics of evolving systems lies in these objects.

\begin{figure}
\centerline{\includegraphics[scale=0.44,clip=]{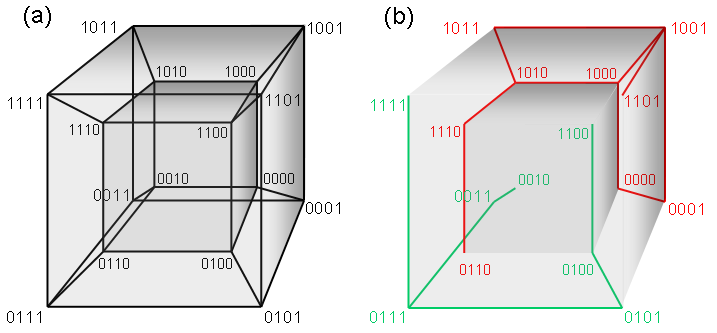}}
\caption{(a) Genotype space for a sequence of length $L=4$ and an alphabet of
$k=2$ letters, $\{0,1\}$. (b) An example of how this space could split into 
two different neutral networks (marked with different colors), each yielding
a different phenotype. 
\label{FGS}
}
\end{figure}

Consider a biological sequence of length $L$. Position $i$ of this sequence can
adopt one out of $k$ variants, that we can think of as letters of an
alphabet. Depending of the type of sequence this alphabet may be formed by the
$k=4$ bases of which DNA or RNA are made of, by the $k=20$ aminoacids that
build up proteins, or even by the $k$ different alleles of gene $i$ of a given
chromosome. The description is similar in any of these instances but we shall
focus e.g.\ on DNA to fix ideas. Every realization of a DNA sequence is a
genotype, and represents a ``point'' in genotype space. There are $4^L$
$L$-long different genotypes; if $L=100$, for instance ---a rather short
sequence, by the way, the size of the genotype space is $4^{100}\approx
10^{60}$, a huge set. Movement across this space proceeds through mutations.
Mutations can be very complicated transformations of a genotype that can even
modify its length, but again, to keep it simple we shall constraint ourselves to
consider only point-like mutations, i.e.\ substitutions of the letter at a
given position by another one of the alphabet. If we now make a graph whose
nodes are all possible sequences and whose links join sequences separated
by a point-like mutation, we have a topological description of a genotype space
(Fig.~\ref{FGS} represents one of these spaces for $L=4$ for an alphabet with
only two letters). Mutations move the sequence from a node of this graph to
one of its $D=3L$ neighbors which differ from it in just one position. 
In general, the genotype space is a regular lattice in a Euclidean space of 
dimension $D=(k-1)L$.

The huge size and high dimensionality of sequence spaces have non-trivial
implications for the distribution of phenotypes in genotype space. Sequences
with the same phenotype have therefore the same fitness, so a sequence can
move across any connected component of the graph corresponding to one
phenotype at no cost in fitness. Figure~\ref{FGS}(b) yields a very simple example 
of sequences that can be accessed without changing the fitness of an individual. 
Note that a single mutation causes no changes if the mutated genome belongs
to the same neutral network than its parental genome. However, in regions where two 
different networks are
close, a point mutation may generate a genome that belongs to a different network,
such that major novelties in phenotype arise.

\begin{figure}[t]
\centerline{\includegraphics[scale=0.3,clip=]{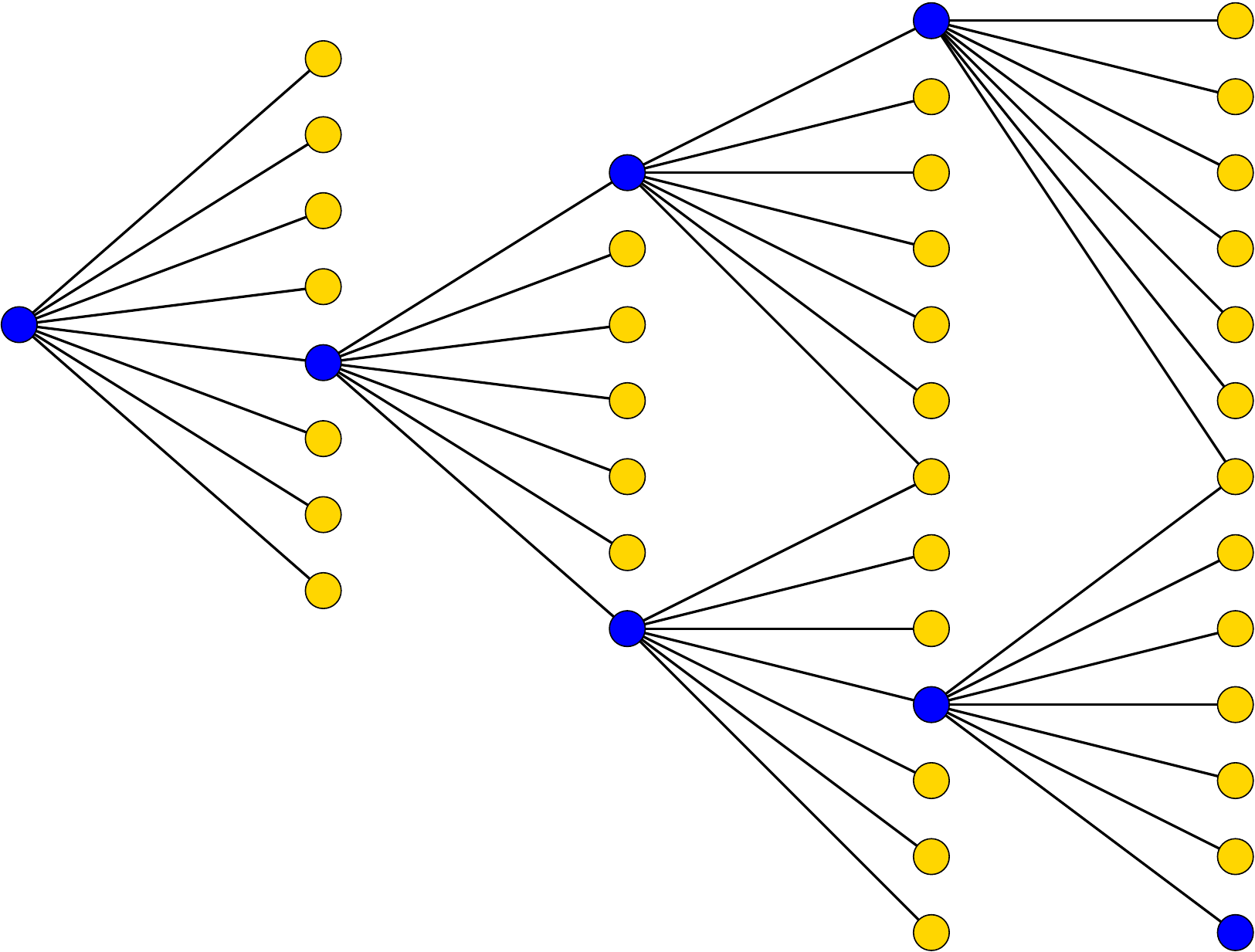}}
\caption{Model of the Russian roulette. Blue nodes belong to the
same phenotype $\Phi$, whereas yellow nodes correspond to different
phenotypes (hence to different fitness values, in principle). If the
fraction $p$ of blue nodes times the number $D$ of links to nearest
neighbors is above $1$, the cluster of blue nodes will extend all over
the network.
\label{FRR}
}
\end{figure}

In order to better understand what these
connected components look like let us consider a simple model in high dimensions
---i.e.\ for genomes which are longer than that of Fig.~\ref{FGS}.
Let us assume
that sequences are randomly and independently assigned to phenotypes, and
let $p$ be the fraction of sequences corresponding to a given phenotype $\Phi$. 
Due to the complexity of the genotype space we can locally regard it as a tree
(see Fig.~\ref{FRR}). Given a node, each of its $D$ neighbors has $D-1$
new neighbors; each of these second neighbors of the first node will have,
in its turn, $D-1$ new neighbors; and so on. Now, because nodes belong to
$\Phi$ randomly and independently of each other, assuming that the first node
belongs to $\Phi$, each second, third, etc., neighbor will also belong to
$\Phi$ with probability $p$. If $(D-1)p>1$, on average every $\Phi$-node
will have another $\Phi$-node among its neighbors, so the set of $\Phi$-nodes
contains a connected cluster with a finite fraction of all the nodes of
the graph. On the contrary, if $(D-1)p<1$ eventually the number
of $\Phi$-nodes will drop to zero, and so the set of $\Phi$-nodes will be
made of ``small'' disconnected clusters. Notice that the critical fraction
of nodes is $p_c\approx D^{-1}$, a very small number in high-dimensional
spaces, so what we have just described is the typical situation.

The picture this provides is very different from that of the standard
fitness landscapes employed in population genetics. Here genotype
spaces should be thought of as a patchwork of different phenotypes,
each patch containing a finite fraction of the total set of nodes, all
of which have the same fitness. Patches are intertwined in very irregular
ways.

\begin{figure}
\centerline{\includegraphics[scale=0.44,clip=]{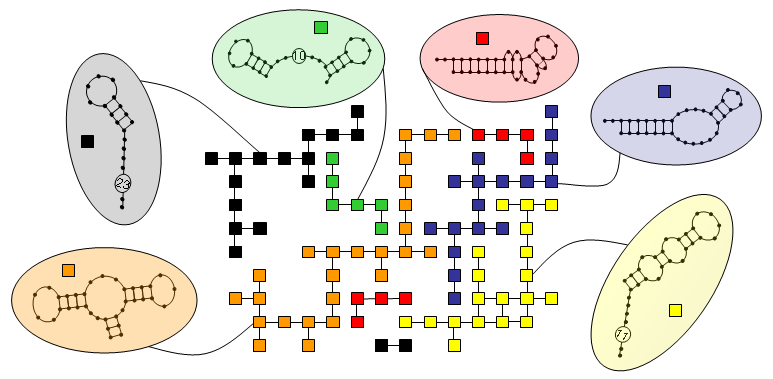}}
\caption{A simple example of the redundant relation between genotype and phenotype. 
Genotypes are represented as squares. Two genotypes (sequences) folding into the same 
secondary structure belong to the same neutral network. Changes in a single nucleotide 
can lead to a complete rearrangement of the folded state, and thus to a significantly 
different phenotype. Typically, the genome space of RNA folding
is such that many different phenotypes can be attained by changing only a few 
positions in the sequence. 
\label{F4} 
}
\end{figure}

Again, RNA folding can give us a quantitative picture of how a neutral network of 
genotypes should look like, and how different networks are interrelated. Suppose that 
one can construct the complete mapping of RNA sequences of a given length into the 
secondary structures they fold into. The genome space would be partitioned into a large 
number of neutral networks, as sketched
in Fig.~\ref{F4}. 
The size of neutral networks varies broadly around an average of $(4/b)^l l^{3/2}$
sequences per network. For example, in the case of sequences of length $l=35$, 
there are around $10^3$ structures (called \emph{common structures)}
which are a thousand-fold 
more frequently obtained from the folding of a randomly chosen sequence than a background 
of millions of other structures that are yielded by few selected sequences \cite{Sti08}.
Interestingly, 
the functional structures found in Nature, though arising from a long and demanding 
selection process through geological time, all belong to the set of common structures. 
The network of genotypes corresponding to common structures traverses the whole space 
of genomes. In practice, thus, a population can contain a huge number of different 
genotypes with identical selective value. Populations can spread in the space of genomes 
without seeing its fitness affected. One important implication of the above is 
accessibility: almost any other possible secondary structure can be accessed with one or 
few changes in the sequence, since networks belonging to different folds have to be 
necessarily close to one or another of the common structures. Systematic measures 
with RNA structures indicate that any common structure 
lies at most $R$ nucleotides apart, with $R \simeq 0.2 l$,
of any other randomly chosen common structure \cite{Gru96}. 

Evidence for the spread of neutral networks throughout the sequence space, and for the
existence of sequences performing different chemical functions (thus having different
phenotypes) that lie just a few nucleotides apart, comes not only from RNA, but also
from empirical results with aptamers and ribozymes. In a revealing experiment,
Schultes and Bartel
\cite{Sch00} discovered close contacts between the neutral networks representing
a class-III self-ligating ribozyme and that of hepatitis$-\delta$ virus self-cleaving
ribozyme. The experiment began with the two original RNA sequences of the
corresponding functional molecules, which had no more than the 25\% similarity
expected by chance. After about 40 moves in genome space,
they located an intersection between the two neutral networks where two sequences
just two nucleotides apart could perform the original functions without a major loss
in fitness. This observation has been repeated in several other systems (see
Ref.~\cite{Sch06} for a review).

An illustration of the relationship between genomes, neutral network spreading and 
phenotypes is represented in Fig.~\ref{F4}. Even in this two-dimensional representation it 
is clear how moving on a neutral network (thus conserving fitness) permits to access 
different phenotypes in a single mutational move. This property might underlie
punctuated equilibrium, explaining the sudden changes in phenotypes observed 
after long periods of stasis \cite{Fon98}. The movement of the population on the 
neutral network, though having effects at the genomic level, does not cause any visible 
change. However, if a better phenotype is encountered through this silent evolution 
behind the curtain, it will be fixed in the population rapidly (due to its advantage 
compared to the previously dominating one) in what will be interpreted as a punctuation 
of the dynamics. Note, however, that the population will then be genomically trapped in a 
position of the neutral network close to the old phenotype. It will take a while until it 
diffuses again on the new network and is able to access different, maybe improved, 
phenotypes. 

Punctuated equilibrium was first defined in relation to the fossil 
record \cite{Eld72}, and yet we have used a simple computational model for RNA 
folding to describe it. The question arises: has this process been observed also at the 
molecular level in natural systems? And the answer is yes. The process of spreading on a 
neutral network followed by a selective sweep when the population discovers a new, fitter
phenotype, plus the subsequent
exploration (again spreading) without phenotypic change to repeat the discovery of
innovation, and so on, has been observed in the yearly dynamics of influenza~A
\cite{Koe06}. This dynamics describes the replacement every $2$ to $8$ years of 
circulating populations (where all individuals share a genetically similar hemaglutinine),
by new populations, different from the previous one (but whose individuals
again share similar sequences).
Hemaglutinine is a protein that determines the antigenic properties of the virus:
continuous changes in this protein permit influenza to escape immunity. This case constitutes
a wonderful example of how relevant it is to use an appropriate genotype-phenotype map to 
understand the co-evolution of pathogens and hosts ---or the adaptation properties of
quasispecies. 

\section{Fitness landscapes and evolution on neutral networks}

In order to understand the complex interplay between the fitness of genomes
(which is determined by the adaptation that they provide to a specific environment)
and the topology of the genome space, different paradigmatic fitness landscapes
have been devised.
Their introduction has been very much conditioned by the interest in 
obtaining analytical results describing the dynamics of quasispecies and other complex 
populations, as well as the characteristics of the process of adaptation and of the 
mutation-selection equilibrium. One of the most popular fitness landscapes is the 
single-peak landscape. Usually, it is assumed that a privileged 
genotype has the largest 
fitness and all the rest have lower fitness, well below that of
the fittest sequence, or even zero. 
The Fujiyama landscape is smoother (also more complex) since it assumes that fitness of 
genotypes decreases with the number of mutations with respect to the fittest type. At 
the other extreme, we find rugged landscapes, among which two prototypical
examples are the random landscape, where each genotype is assigned a randomly 
and independently chosen fitness value, or Kauffman's $NK$-landscapes, in which
each of the $N$ genes of a sequence contributes additively to the fitness of the
genome, but its fitness value 
results from its epistatic interactions (typically random) with $K$ other genes.
There is not much in between, where one would guess that realistic 
landscapes should lie.

\begin{figure}[t]
\centerline{\includegraphics[scale=0.35,clip=]{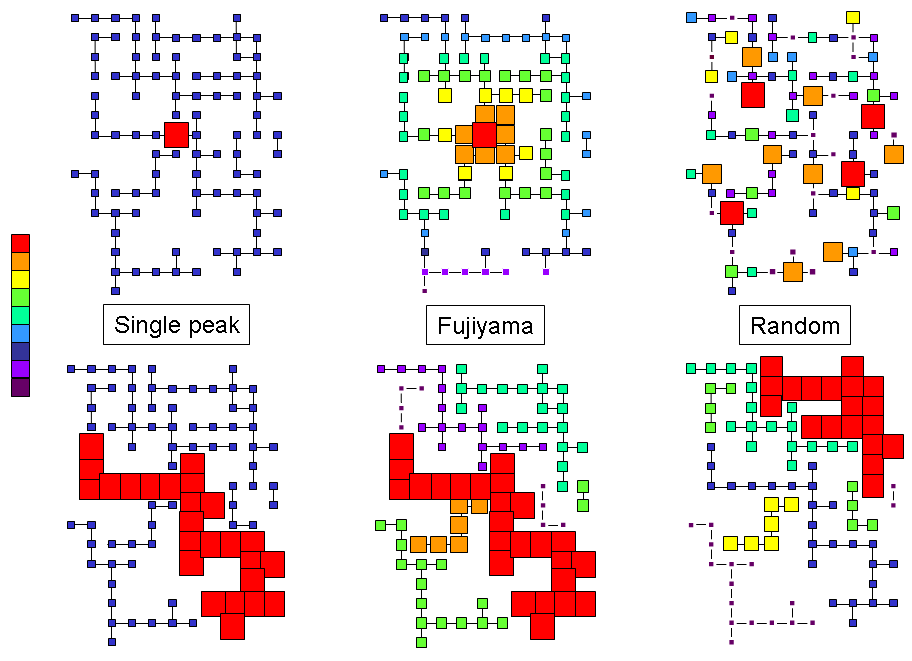}}
\caption{Schematic representation of three different fitness landscapes
(as indicated) describing the differences between a genotype-based fitness
and a phenotype-based fitness.
Fitness values are proportional to the size of the nodes, and absent nodes
are assumed to be non-viable sequences (zero fitness).
The single peak landscape privileges one particular genome (above) or
one particular phenotype (below). 
The Fujiyama landscape assigns maximum fitness
to one sequence (above) or to one phenotype (below). Fitness decreases as
the distance from each sequence or from each phenotype to the 
optimum increases. Note that this rule yields a smooth landscape 
only in sequence space, 
since phenotypes change much more abruptly.
In the latter case,
it resembles a random landscape (last column, below).
The landscape defined by RNA folding shares many properties with random
landscapes. A random assignation of fitness in genome space (last column,
above) leads to a truly decorrelated landscape. 
\label{F5} 
}
\end{figure}

But, according to
the picture we have just drawn, fitness landscapes should incorporate the
high redundancy observed in biological sequences. Now we know that genotypes
organize themselves into regions of common phenotypes, which therefore have
constant fitness and which spread all over the genome space,
forming so-called \emph{neutral networks.} We can then try to figure out what the
prototypical fitness landscapes should look like when these neutral networks of
common phenotypes are taken into account. This is what Fig.~\ref{F5} summarizes.
The top row of that figure sketches a representation of the single peak, the
Fujijama, and the random landscapes, as referred to single genotypes. The single
peak exhibits a single point of high fitness in a sea of points of lower or zero
fitness. In the Fujijama landscape, points decrease in fitness as the get away
from the optimum sequence. In the random landscape points have random fitness,
independently of each other.
The lower row of Fig.~\ref{F5} shows the phenotype counterparts of these three
archetypes. Points are arranged into networks of constant fitness
(equal phenotype), so the single peak now shows one of this networks with
high fitness surrounded by other networks of low fitness and by non-viable genotypes
(zero fitness). The Fujijama landscape is now defined in terms of distance 
between phenotypes, producing a landscape not quite distinguishable from
what a random landscape now looks like. 

In order to describe evolution in these new fitness landscapes we need
new mathematical tools to deal with neutral networks \cite{Rei02}. Neutral networks
can be described through a connectivity matrix $C$, whose elements
are $c_{ij}=1$ if genotypes $i$ and $j$ are mutually accessible and $0$ otherwise.
Evolution and adaptation, understood as a process of search and fixation of fitter
phenotypes, is conditioned by the topology of these connectivity matrices
and by the relationships between them, understood as objects defined in the space of
genomes. There are a number of results that relate the topology of
those graphs with the equilibrium states of populations and the dynamics of adaptation
on the neutral network. It has been shown that the distribution of a population
evolving (i.e. replicating and mutating) on a
neutral network is solely determined by the topological properties of $C$, and
given by its principal eigenvector. In that configuration the population has evolved
mutational robustness, since it is located in a region of the neutral network where the
connectivity is as large as possible (thus where mutations affect as less as possible
the current phenotype) \cite{Nim99}. This maximal connectivity equals the
spectral radius of $C$. Equilibrium properties are thus well described once $C$ is
known.

The dynamics of adaptation on neutral networks are more difficult to fully
quantify because, in principle, all eigenvalues of the matrix $C$ intervene in the transient
towards equilibrium. In addition, the time required to reach the equilibrium configuration
depends on the initial condition: it might differ in orders of magnitude (in
units of generations) if the population enters the network through a particular node
---as in the case of influenza A--- or if all genomes are equally represented ---as in
{\it in vitro} experiments that begin with a large population of random sequences. It
has been shown that time to equilibrium is inversely proportional to the mutation rate,
such that homogeneous populations
(low mutation rates) will have it difficult to develop high mutational
robustness. In very general conditions, the dominant term in the time to equilibrium is
proportional to the ratio between the second largest and the largest eigenvalue of
$C$ \cite{Agu09}. $C-$matrices are highly sparse, symmetric matrices
for which it seems
likely to develop approximations that could yield their two largest eigenvalues as a
function of the average connectivity, for instance. To this end, the analysis of neutral
networks could be performed in the limit of infinite size, given their exponentially fast
growth in size with the sequence length.

Finally, an essential ingredient in the evolutionary process is randomness, and not
only in relation to genetic drift. Random fluctuations play a main role in the
searching process. Too low a variability in a population might even
completely block adaptation.
For example, the quantity that determines whether a population will be able to
attain the region of maximal neutrality in finite time is the product of the population
size times the mutation rate \cite{Nim99}. Higher adaptability can be reached by means of
a large population or through a large mutation rate. Overly small or homogeneous populations
might get trapped in suboptimal configurations analogous to the metastable states
observed in disordered systems.

A deeper knowledge of the topological properties of neutral networks and their mutual
relationship in sequence space should lead to more realistic dynamical models for the
evolution of populations. Provided one could characterize the fitness landscape, the
probability of changing from one phenotype to another would be described through a matrix
of transitions $M=(m_{ij})$ between states, with $m_{ij}\ge 0$.
This is actually a common formal
framework to study population dynamics \cite{Bly07}. Matrices $M$ are stochastic,
i.e.\ $\sum_j m_{ij}=1$ and thus define a homogeneous Markov chain.
A full knowledge of the dynamics of the system amounts to
knowing the eigenvalue spectrum of $M$.

\section{Conclusions}

The process of adaptation is not strongly relying on happy coincidences. The 
existence of huge and extensive neutral networks permits systematic explorations of the
space of possible functions without paying high fitness costs ---a practical way to find 
out viable pieces later assembled to form complex individuals. Our current 
understanding of the relationship between genotype and
phenotype clearly hints at the fact that even an evolutionary process restricted in the
amount of change it can produce at the genomic level is not necessarily restricted in the 
amount of change it can cause at the phenotypic level. Further, it seems plausible that all 
possible phenotypes are sufficiently close to each other, such that it is not necessary to 
explore all the space of genotypes to find the optimal phenotype. While a genotype might 
be the needle in a haystack, you can't help but stumble upon the phenotype. 

This picture of a space of genomes where neutral networks corresponding to common
functions are vastly extended and deeply interwoven has important
implications in the way we understand and model the evolutionary process. Fast mutating
populations, as RNA viruses, are able to spread rapidly and find new adaptive solutions
thanks to the sustained generation of new viral types and the costless drift through large
regions of genome space. Due to their relatively short genomes and the continuous 
accumulation of new mutations, it is very difficult (impossible in many cases) to trace
the ancestry of extant viruses. Thus, viral phylogeny is located in evolutionary time, and 
the signal that speaks for its origins becomes increasingly weaker as we move backwards, 
until it is eventually lost. As a result, there is an on-going controversy on the origin of 
viruses, on their being a product of the post-cellular era or the remnants of an ancient, 
pre-cellular RNA world. High mutation rates have been a successful strategy in their case, 
allowing the perpetual exploration of new genomic regions and thus escaping the attack of 
their hosts' defenses.

But when we come to talk about Life on Earth, with all the amazing complexity and 
diversity of organisms formed at least by one cell, it turns out that their common origins
can be unequivocally identified. The phylogeny reconstructed through ribosomal units, single
genes or whole genomes of living organisms clearly reveals the existence of LUCA, our Last 
Universal Common Ancestor, some $3.5$ billion years ago. Is thus life on Earth resting on 
a frozen accident, that is the precise genomic pieces that formed LUCA? In the light of the 
above, we should answer ``no''. The first genomes could have occupied far-away places in the 
space of genomes and, still, it is highly improbable that functional life would look nowadays 
very different from the solutions (the phenotypes) we see all around us. 





\begin{acknowledgments}
The authors acknowledge support from the Spanish Ministerio de Educaci\'on y
Ciencia under projects FIS2008-05273 and MOSAICO, and from DGUI of the Comunidad de
Madrid under the R\& D program of activities MODELICO-CM/S2009ESP-1691.
\end{acknowledgments}





\end{article}








\end{document}